\RequirePackage{arydshln}
\documentclass[aps,twocolumn,nofootinbib,superscriptaddress,preprintnumbers,pra,10pt,floatfix]{revtex4-1}

\usepackage[utf8]{inputenc}
\usepackage{float}
\usepackage{colortbl}
\usepackage{multirow}
\usepackage{amsmath,amssymb,amsfonts, bm,bbm,slashed}
\usepackage{dsfont} 
\usepackage{hyperref}
\usepackage{graphicx}
\usepackage{enumitem}
\usepackage{arydshln}
\usepackage{mathtools}
\usepackage{bbold}
\usepackage{braket}
\usepackage{soul}

\makeatletter
\g@addto@macro\bfseries{\boldmath}
\makeatother

\newcommand{\be} {\begin{equation}}
\newcommand{\ee} {\end{equation}}
\newcommand{\bea} {\begin{eqnarray}}
\newcommand{\eea} {\end{eqnarray}}
\newcommand{\no} {\nonumber}
\newcommand{\ba} {\begin{array}}
\newcommand{\ea} {\end{array}}
\newcommand{\gsim}{\lower.7ex\hbox{$\;\stackrel{\textstyle>}{\sim}\;$}}
\newcommand{\lsim}{\lower.7ex\hbox{$\;\stackrel{\textstyle<}{\sim}\;$}}
\renewcommand{\Re}{{\rm Re}}

\usepackage{lipsum}
\usepackage[dvipsnames]{xcolor}

\begin{document}

\preprint{ZU-TH-25/20}

\title{Optimized Lepton Universality Tests in $B\to V \ell\bar{\nu}$ decays}
 
\author{Gino Isidori}
\email{isidori@physik.uzh.ch}
\affiliation{Physik-Institut, Universit\"at Z\"urich, CH-8057 Z\"urich, Switzerland}
\author{Olcyr Sumensari}
\email{olcyr.sumensari@physik.uzh.ch}
\affiliation{Physik-Institut, Universit\"at Z\"urich, CH-8057 Z\"urich, Switzerland}

\begin{abstract}
We propose improved Lepton Flavor Universality (LFU) ratios in semileptonic $P\to V \ell \bar{\nu}$ decays, 
when comparing $\mu$ and $\tau$ modes,  that minimize the theoretical form-factor uncertainties. 
These optimized ratios are obtained with simple 
cuts or reweighting of the dilepton mass distributions, which imply a minimum loss of signal on the rare tauonic modes
while maximizing the cancellation of theoretical uncertainties among the two modes.  
We illustrate the usefulness of these observables in $B_c\to J/\psi$, $B_c\to \psi(2S)$, $B\to D^\ast$ and $B_s\to D_s^\ast$ transitions,
showing that in all cases we can reach $\mathcal{O}(1\%)$ uncertainties on the SM predictions of the 
improved LFU ratios employing conservative form-factor uncertainties. 
\end{abstract}

\maketitle

\allowdisplaybreaks


\section{Introduction}\label{sec:intro}

The hints of Lepton Flavor Universality (LFU) violation in charged-current semi-leptonic $b\to c \ell \nu$
decays~\cite{Lees:2013uzd,Huschle:2015rga,Aaij:2015yra,Hirose:2016wfn,Aaij:2017deq,Hirose:2017dxl,Belle:2019rba}, as well as in  $b\to s \ell\ell$ transitions~\cite{Aaij:2014ora,Aaij:2017vbb,Aaij:2019wad,Abdesselam:2019wac,Abdesselam:2019lab}, represent one of the most 
fascinating challenges in particle physics. 
Recent data confirm numerous discrepancies from the Standard Model (SM) predictions in both sectors.
 At present there is not a single measurement with a high statistical significance,  but
the global picture is very consistent. These hints indicate a LFU violation of short-distance origin, 
encoded in the four-fermion semileptonic interaction.

In this letter we focus the attention on the LFU tests in  $b\to c \ell \nu$ transitions, which  so far is the sector with the lowest 
statistical significance~\cite{Gambino:2020jvv}. Beside improving the measurements of  $R_D$ and $R_{D^*}$, it would be desirable 
to add more observables able to probe the same underlying partonic transition. A first step in this direction has been 
undertaken by the LHCb collaboration~\cite{Aaij:2017tyk} with the measurement of the $R_{J/\psi}$ ratio in $B_c$-meson decays, 
obtaining
\be
R_{J/\psi} \doteq \dfrac{\mathcal{B}(B_c \to J/\psi \tau \bar{\nu})}{\mathcal{B}(B_c\to J/\psi \mu \bar{\nu})} = 
0.71 \pm 0.17_{\rm stat}  \pm 0.18_{\rm syst}~.
\label{eq:RJpsi}
\ee
This result has to be compared with a SM prediction ranging between 0.25 and 0.28~\cite{Ivanov:2006ni,Dutta:2017xmj,Watanabe:2017mip,Tran:2018kuv,Leljak:2019eyw,Cohen:2019zev}, with an error estimated to be around 10\% in~\cite{Cohen:2019zev}.
The large experimental error, as well as the sizable theoretical uncertainty in the SM prediction, do not allow
us to draw significant conclusions from this result at present.  

The source of the SM error on $R_{J/\psi}$  and, partially, also of the 
systematic error in the experimental result, is the poor knowledge of the $B_c \to J/\psi$ hadronic form factors.
The knowledge of the latter is expected to improve soon thanks to lattice QCD calculations~\cite{Aoki:2019cca}.
However, it is desirable to develop alternative methods to reduce this source of uncertainty. 
The purpose of this letter is to propose improved LFU ratios on  $B_c \to J/\psi \ell \bar{\nu}$  and, more generally,
any $P\to V\ell\bar{\nu}$ decay suffering from large form-factor uncertainties, which would allow us to minimize the 
 error on the corresponding SM predictions.  
As we will show, in these channels we can reduce the theory error on appropriate LFU ratios -- at fixed form-factor uncertainty --
taking into account that: i)~the only intrinsic theory error (i.e.~the uncertainty associated to the non-universal part of the amplitude) 
is the one induced by the scalar form factor; ii)~the scalar form factor generates a subleading contribution to the 
decay rate that vanishes at large dilepton invariant mass.

\section{$P\to V\ell\bar{\nu}$ decays}\label{sec:formulas}

\subsection{SM description}

We consider a generic process of the type $P\to V\ell \bar{\nu}$, based on the underlying partonic 
transition $b\to c \ell \bar{\nu}$, where $P$ and $V$ denote a pseudoscalar and vector meson. Within the SM, the branching fraction for this process can be written as~\cite{Korner:1989qb},
\begin{align}
\label{eq:BR-def}
\dfrac{\mathrm{d}\mathcal{B}}{\mathrm{d}q^2}(P\to V\ell \bar{\nu}) = \Phi(q^2)\, \omega_\ell(q^2)\,\big{[}H_V^2 + \left(H_S^\ell\right)^2\big{]}\,,
\end{align}

\noindent where $q^2=(p_\ell +p_\nu)^2$ is the dilepton invariant mass squared.
The phase-space factors are 
\begin{align}
\Phi(q^2) &\doteq \mathcal{B}_0\, q^2 \, \sqrt{\lambda_V(q^2)}\,,\\[0.5em]
\omega_\ell (q^2) &\doteq\left(1-\dfrac{m_\ell^2}{q^2}\right)^2\left(1+\dfrac{m_\ell^2}{2q^2}\right)\,,\\[0.5em]
\lambda_V(q^2) &\doteq \left[q^2-(M-m)^2\right]\left[q^2-(M+m)^2\right]\,,
\end{align}

\noindent 
where $M$ and $m$ denote the $P$ and $V$ meson masses, respectively,
$\mathcal{B}_0=\tau_P\,G_F^2 |V_{cb}|^2/(192 \pi^3 M^3)$,  and $\tau_P$ stands for the lifetime of the $P$ meson. The hadronic matrix elements are fully encapsulated in the helicity amplitudes  $H_V$ and $H_S^\ell$, 
\begin{align}
\label{eq:HV}
H_V^2 &= H_{V^+}^2+H_{V^-}^2+H_{V^0}^2\,, \\[0.4em]
\label{eq:HS}
\left(H_{S}^\ell\right)^2 &= \dfrac{3m_\ell^2}{m_\ell^2+2q^2}H_{V^t}^2\,,
\end{align}

\noindent with
\begin{align}
\label{eq:HVpm}
H_{V^\pm} &= (M+m)\,A_1(q^2)\mp \dfrac{\sqrt{\lambda_V (q^2)}}{M+m}\,V(q^2)\,,\\[0.4em]
\label{eq:HV0}
H_{V^0} &= \dfrac{M+m}{2 m \sqrt{q^2}}\bigg{[}-(M^2-m^2-q^2)\,A_1(q^2)\\
&\qquad\qquad\quad+\dfrac{\lambda_V(q^2)}{(M+m)^2}\,A_2(q^2)\bigg{]}\,,\nonumber\\[0.4em]
\label{eq:HVt}
H_{V^t} &= -\sqrt{\frac{\lambda_V(q^2)}{q^2}}\,A_0(q^2)\,,
\end{align}
\noindent where $V$, $A_{0}$, $A_{1}$ and $A_{2}$ are the $P\to V$ form-factors collected in Appendix \ref{app:form-factors}, and the polarizazion vectors are defined as in Ref.~\cite{Becirevic:2019tpx}.

\subsection{LFU ratios}
\label{ssec:LFU}

The usual LFU ratios are defined as the ratio of the inclusive rates (or branching fractions) 
for different lepton modes,
\be
\label{eq:RV-normal}
R_V  \doteq   \dfrac{ \Gamma( P\to V\tau \bar{\nu} ) }{  \Gamma( P\to V\mu \bar{\nu} ) }~,
\ee
such that
\be 
R_V^{\rm SM}  = \dfrac{\displaystyle\int_{m_\tau^2}^{q^2_\mathrm{max}} \mathrm{d}q^2\, \Phi(q^2)\, \omega_\tau(q^2)\,\big{[}H_V^2 + \left(H_S^\tau\right)^2\big{]}}{\displaystyle\int_{m_\mu^2}^{q^2_\mathrm{max}} \mathrm{d}q^2\, \Phi(q^2)\, \omega_\mu(q^2)\,\big{[}H_V^2 \big{]}}\,, \quad 
\label{eq:RV-normal-SM}
\ee
where $q^2_\mathrm{max}=(M-m)^2$. In  (\ref{eq:RV-normal-SM}) 
we have neglected the scalar helicity amplitude in the denominator since it is suppressed by the muon mass. 
Within the SM, $R_V$  is not equal to unity because of three different effects:
\begin{itemize}
\item[(i)] The different integration ranges in the numerator and denominator.
\item[(ii)] The different weights $\omega_\ell$ of the $H_V$ contributions for $\mu$ and $\tau$ modes.
\item[(iii)] The scalar contribution $H_S^\ell$, which is numerically relevant only for the $\tau$ mode.
\end{itemize}
Due to these three effects, there is only  a partial cancellation of the hadronic uncertainties in $R_V$.
In particular, only the overall normalization error on the leading  $H_V$ term cancels between numerator 
and denominator, but not the error associated to its $q^2$ dependence. 
On the other hand, it is clear that the uncertainties associated to the effects~(i) and~(ii) can be eliminated 
if, in addition to the total rate, also the $q^2$ spectrum were experimentally accessible. 
The lepton mass dependence induced by~(i) and~(ii) is indeed a know function of $q^2$. 
 The only irreducible error is the one associated to (iii),  
 which can also be reduced via a differential $q^2$ measurement noting that the relative contribution of $H_S^\ell$ to the decay rate 
decreases at large $q^2$. These observations are at the basis of the improved LFU observables
that we introduce below.


\subsection{Improved LFU ratios}

The first improvement with respect to the usual definition is to use the same integration range in the numerator and denominator,

\bea
\label{eq:RV-cut}
&& R_V^{\mathrm{cut}} \big{(}q^2_\mathrm{min}\big{)} \doteq \dfrac{\displaystyle\int_{q^2_\mathrm{min}}^{q^2_\mathrm{max}}\mathrm{d}q^2\,\dfrac{\mathrm{d}\Gamma}{\mathrm{d}q^2}(P\to V\tau \bar{\nu})}{\displaystyle\int_{q^2_\mathrm{min}}^{q^2_\mathrm{max}}\mathrm{d}q^2\,
\frac{\mathrm{d}\Gamma}{\mathrm{d}q^2}(P\to V\mu\bar{\nu})}~,
\eea

\noindent where $q^2_\mathrm{min} \geq m_\tau^2$. This simple modification allows to eliminate the 
source of error~(i) listed above. More precisely, using the same integration range we get rid of the uncertainty on the 
muon mode arising from the (non-interesting) kinematical region where we cannot compare it to the tau mode.
This point was noted first in~\cite{Freytsis:2015qca,Bernlochner:2016bci}, where the measurement of   $\widetilde R_V\doteq R_V^{\mathrm{cut}}(m_\tau^2)$
was proposed. We note here that setting $q^2_\mathrm{min}> m_\tau^2$ can be more 
convenient since it allow us to partially address also the points (ii) and~(iii) listed above, 
at the price of a (minor) increase of the statistical error on the measurement. 
Indeed, for  large $q^2$  the weights $\omega_\ell (q^2)$ converge to unity,  for all lepton flavors, and $H_S^\ell$ becomes negligible.

Beside choosing a common phase space region for numerator and denominator in the ratio, a further reduction of the 
theory error can be obtained by a suitable $q^2$-dependent reweighting of the light-lepton rate. More precisely, we propose to 
measure the following optimized observable:
\begin{equation}
\resizebox{0.425\textwidth}{!}{$
R_V^{\mathrm{opt}}  \big{(}q^2_\mathrm{min}\big{)}\doteq \dfrac{\displaystyle\int_{q^2_\mathrm{min}}^{q^2_\mathrm{max}}\mathrm{d}q^2\,\dfrac{\mathrm{d}\Gamma}{\mathrm{d}q^2}(P\to V\tau \bar{\nu})}{\displaystyle\int_{q^2_\mathrm{min}}^{q^2_\mathrm{max}}\mathrm{d}q^2\,\left[\frac{\omega_\tau(q^2)}{\omega_\mu(q^2)}\right]\, \frac{\mathrm{d}\Gamma}{\mathrm{d}q^2}(P\to V\mu\bar{\nu})}\,. $}
\end{equation}
By construction, the reweighting of the muon rate in $R_V^{\mathrm{opt}}$ is such that the leading $H_V$ terms appear 
with the same coefficient, for any $q^2$ bin, in both numerator and denominator. The corresponding SM prediction is
\begin{equation}
\label{eq:RV-opt}
\resizebox{0.425\textwidth}{!}{$
\left. R_V^{\mathrm{opt}}  \big{(}q^2_\mathrm{min}\big{)} \right|_{\rm SM} 
=\dfrac{\displaystyle\int_{q^2_\mathrm{min}}^{q^2_\mathrm{max}} \mathrm{d}q^2\, \Phi(q^2)\, \omega_\tau(q^2)\,\big{[}H_V^2 + \left(H_S^\tau\right)^2\big{]}}{\displaystyle\int_{q^2_\mathrm{min}}^{q^2_\mathrm{max}} \mathrm{d}q^2\, \Phi(q^2)\, \omega_\tau(q^2)\,\big{[}H_V^2 \big{]}}\,$}~,
\end{equation}
addressing completely the points (i) and  (ii) in section~\ref{ssec:LFU}.
As for $R_V^{\mathrm{cut}} \big{(}q^2_\mathrm{min}\big{)}$, the point~(iii) can be partially addressed, 
at the price of an increase of the statistical error, setting $q^2_\mathrm{min} > m_\tau^2$. 
As expected by a theoretically clean LFU ratio,  $R_V^{\mathrm{opt}}(q^2_\mathrm{min})$ is predicted to be 
close to unity  in absence of non-standard sources of LFU violations and, as we will demonstrate below, its SM theoretical error is proportional to 
$|R_V^{\mathrm{opt}}(q^2_\mathrm{min})  -1|$. 

\subsection{Theory uncertainty estimation}

To compare the theoretical uncertainty in the SM predictions for  $R_V^\mathrm{cut}$ and $R_V^\mathrm{opt}$, we use a simplified notation for the $q^2$-integral,
\begin{align}
\label{eq:int-simple}
\big{\langle} f\big{\rangle}_\ell \doteq \dfrac{\displaystyle\int \mathrm{d}q^2\, \Phi(q^2)\, \omega_\ell(q^2) \, f(q^2)}{\displaystyle\int \mathrm{d}q^2\, \Phi(q^2)\, \omega_\ell(q^2)}\,,
\end{align}
where $f(q^2)$ is a generic function, and the same integration ranges $q^2 \in (q^2_\mathrm{min},q^2_\mathrm{max})$ in the numerator and denominator are understood. We expand the square of the helicity functions around their central values,
\begin{align}
\begin{split}
H^2_V &\to H^2_V+\delta H^2_V\,,\\[0.4em]
(H_S^\tau)^2 &\to (H_S^\tau)^2 +\delta (H_S^\tau)^2\,.
\end{split}
\end{align}
\noindent The relative error on $R_V^\mathrm{cut}$ induced by $(\delta H_S^\tau)^2$ and $\delta H^2_V$,
expanding to first order,  reads
\bea
\label{eq:Rcut-error}
\left. \dfrac{\delta R_V^\mathrm{cut}}{ R_V^\mathrm{cut} } \right|_S & = &  \dfrac{ 1}{  R_V^\mathrm{cut} } \dfrac{   \big{\langle} \delta (H_S^\tau)^2\big{\rangle}_\tau }{\big{\langle}H_V^2 \big{\rangle}_\mu}\,,  
\no \\
\left. \dfrac{\delta R_V^\mathrm{cut}}{ R_V^\mathrm{cut} } \right|_V  & = &  \dfrac{ 1}{  R_V^\mathrm{cut} }  
\dfrac{   \big{\langle} \delta H_V^2\big{\rangle}_\tau }{\big{\langle}H_V^2 \big{\rangle}_\mu}  
- \dfrac{   \big{\langle}  \delta H_V^2 \big{\rangle}_\mu }{\big{\langle} H_V^2 \big{\rangle}_\mu}\,.
\eea
The total uncertainty is obtained by combining these two terms in quadrature and accounting for possible correlations among them.  
The choice of the same integration region for numerator and denominator implies a cancellation of the two 
$\delta H_V^2$ terms in (\ref{eq:Rcut-error}) which is not exact, but it improves for large $q^2_\mathrm{min}$ where 
$\langle \delta H_V^2 \rangle_\mu \to \langle \delta H_V^2 \rangle_\tau$.

Performing the same expansion on the optimized LFU ratio in Eq.~\eqref{eq:RV-opt}
we obtain
\bea
\label{eq:Ropt-error}
\left. \dfrac{\delta R_V^\mathrm{opt}}{ R_V^\mathrm{opt} } \right|_S & = &  \dfrac{ 1}{  R_V^\mathrm{opt} } \dfrac{   \big{\langle} \delta (H_S^\tau)^2\big{\rangle}_\tau }{\big{\langle}H_V^2 \big{\rangle}_\tau}  = \dfrac{ R_V^\mathrm{opt} -1 }{  R_V^\mathrm{opt} }  
 \dfrac{   \big{\langle} \delta (H_S^\tau)^2\big{\rangle}_\tau }{\big{\langle} (H_S^\tau)^2 \big{\rangle}_\tau}\,, 
\no \\
\left. \dfrac{\delta R_V^\mathrm{opt}}{ R_V^\mathrm{opt} } \right|_V  & = &  \dfrac{ 1 -  R_V^\mathrm{opt}  }{  R_V^\mathrm{opt} }  
 \dfrac{   \big{\langle}  \delta H_V^2 \big{\rangle}_\tau }{\big{\langle} H_V^2 \big{\rangle}_\tau}\,.
\eea
As can be seen, in this case we necessarily have an error proportional to  $|1 -  R_V^\mathrm{opt}|$, i.e.~an error 
proportional to the effective small breaking of LFU implied by the non-vanishing $H^\tau_S$ amplitude.
This is the minimum error one can expect.

\section{improved LFU ratios in specific channels}

In the following we illustrate the usefulness of $R_V^\mathrm{cut}(q^2_{\rm min})$
and $R_V^\mathrm{opt}(q^2_{\rm min})$ with concrete examples in selected decay modes, 
with conservative assumptions on form factor errors.

\

\paragraph{\underline{$B_c\to J/\psi \ell \bar{\nu}$.}} 

\begin{figure}[t]
  \centering
    \includegraphics[width=.97\columnwidth]{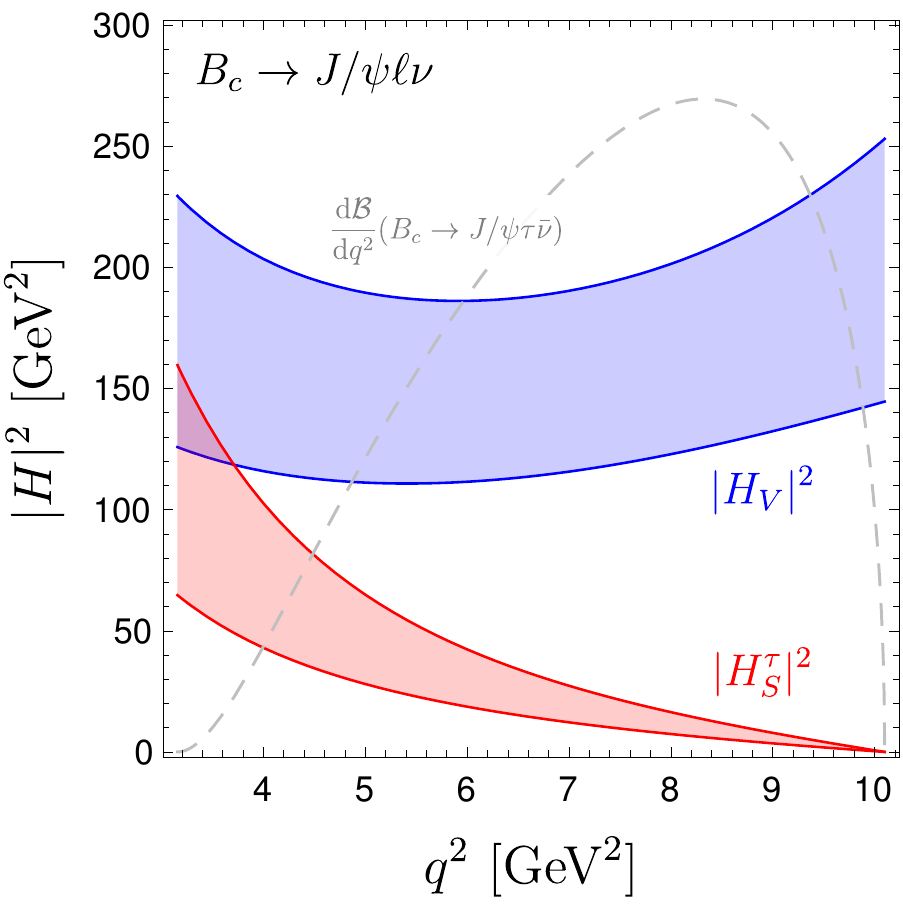}
  \caption{ \sl \small $B_c\to J/\psi$ helicity amplitudes as a function of $q^2$, using the form-factors from Ref.~\cite{Cohen:2019zev}
  The bands denote the $1\sigma$ region. The gray dashed line denotes the differential $q^2$ distribution of $B_c\to J/\psi \tau \bar{\nu}$ in  arbitrary units.}
  \label{fig:Bc-JPsi-helicity}
\end{figure}

Using the  $B_c\to J/\psi$ form factors from Ref.~\cite{Cohen:2019zev} we obtain the bands shown in 
Fig.~\ref{fig:Bc-JPsi-helicity} for the $B_c\to J/\psi \ell \bar{\nu}$ helicity amplitudes.
 As can be seen, the errors are quite large. However, as anticipated, the contribution of 
$H^\tau_S$ vanishes for large $q^2 \to q^2_{\rm max}$.

The standard definition of 
$R_{J/\psi}$ in (\ref{eq:RJpsi}) leads to a $\approx$~10\% error: 
$R_{J/\psi}^\mathrm{SM}=0.25(3)$~\cite{Cohen:2019zev}.
Using the same form factors,  the corresponding predictions for $R_{J/\psi}^\mathrm{cut}$ and $R_{J/\psi}^\mathrm{opt}$, for different values of $q^2_\mathrm{min}$, are shown in Table~\ref{tab:RJPsi}:
setting $q^2_\mathrm{min}=m_\tau^2$ 
the error drops to less than  $6\%$ and $4\%$ for $R_{J/\psi}^\mathrm{cut}$
and $R_{J/\psi}^\mathrm{opt}$, respectively; the error further drops to about $2\%$ in both cases (i.e.~to about 1/5 of the error on $R_{J/\psi}$) 
setting $q^2_\mathrm{min}=7~{\rm GeV}^2$. As can be seen by the grey line in Fig.~\ref{fig:Bc-JPsi-helicity},
a lower  cut at $7~{\rm GeV}^2$ retains
about $85\%$ of the $B_c\to J/\psi \tau \bar{\nu}$ rate, hence the corresponding increase of statistical error is marginal
compared to the drastic reduction of the theory error.

\begin{figure*}[t]
  \centering
    \includegraphics[width=0.47\textwidth]{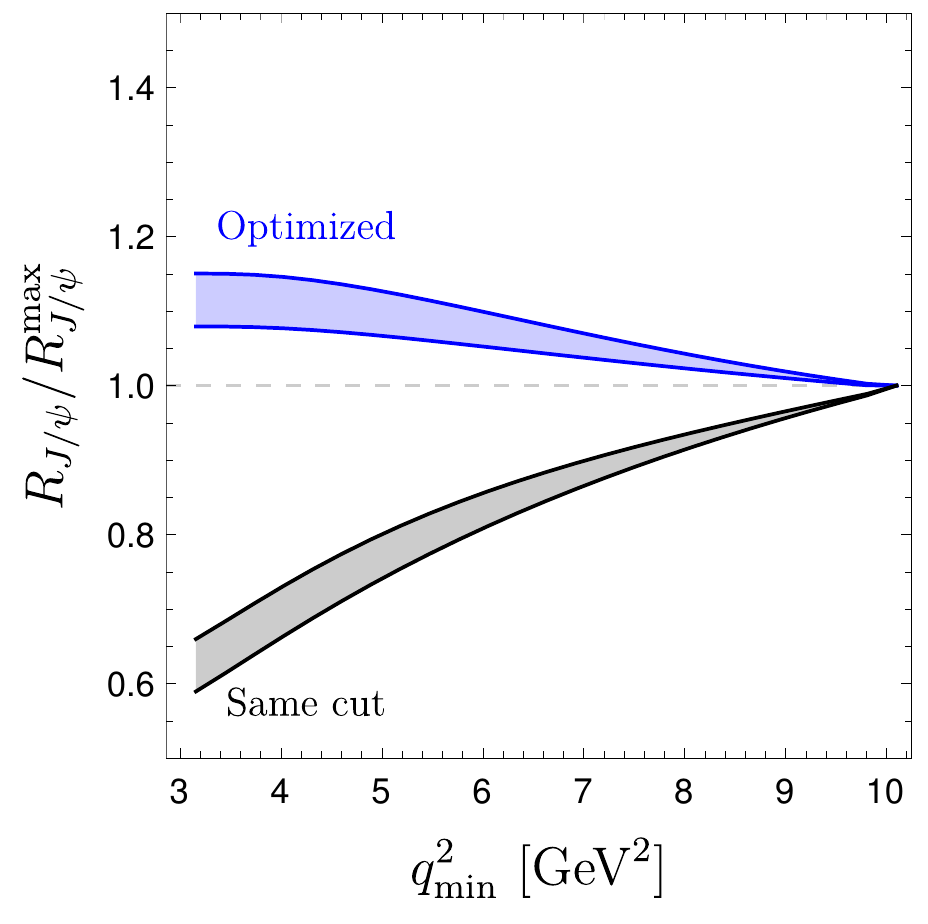}~\includegraphics[width=0.48\textwidth]{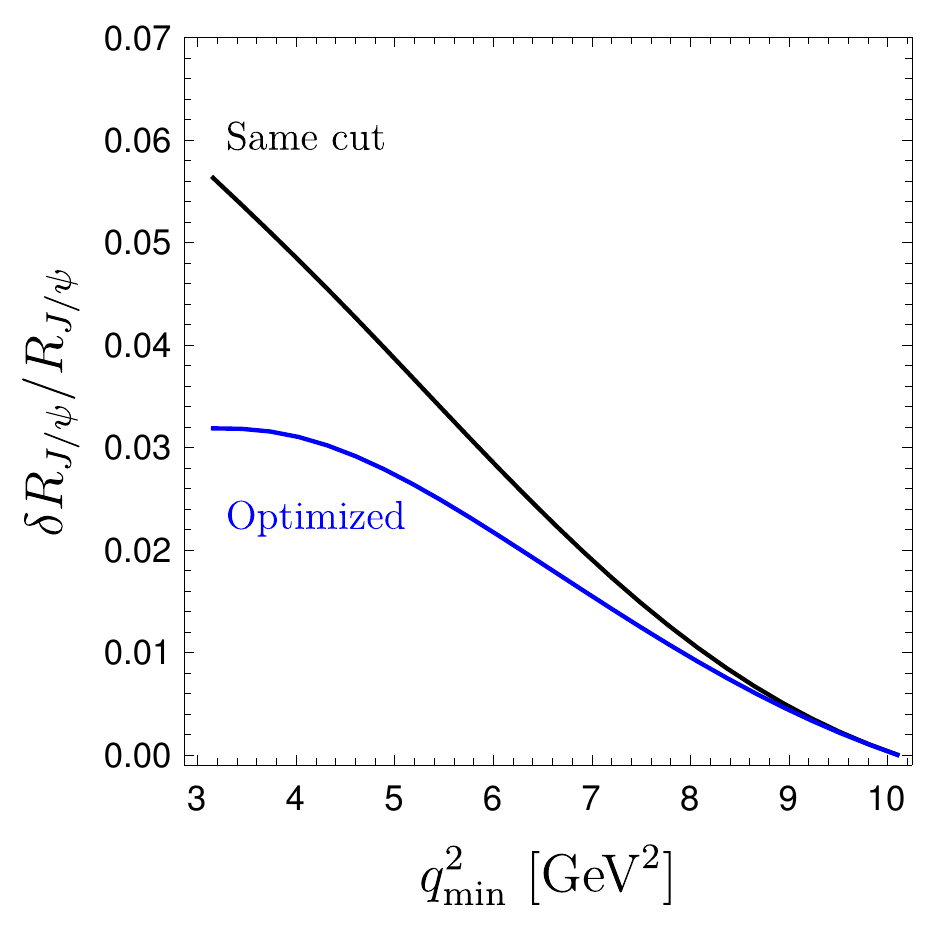}
  \caption{ \sl \small SM predictions for $R_{J/\psi}^\mathrm{cut}$ (black) and $R_{J/\psi}^\mathrm{opt}$ (blue), with $1\sigma$ error band,
 as a function of $q_\mathrm{min}^2$  (left panel). These ratios are normalized by the values at $q^2_\mathrm{max}$, namely $R_{J/\psi}^\mathrm{cut}(q^2_\mathrm{max})=0.549$ and $R_{J/\psi}^\mathrm{opt}(q^2_\mathrm{max})=1$. The comparison of the relative theoretical uncertainty is shown in the right panel.
  }
  \label{fig:Bc-JPsi-prediction}
\end{figure*}

\begin{table}[t]
\renewcommand{\arraystretch}{1.8}
\centering
\begin{tabular}{|c|ccc|c|}
\hline 
$\,q^2_\mathrm{min}\,$ & $m_\tau^2$ & $\,5~\mathrm{GeV}^2\,$  & $\,7~\mathrm{GeV}^2\,$  & form factors\\ \hline\hline
$R_{J/\psi}^\mathrm{\,cut}$  & $0.34(2)$ & $0.42(2)$ & $0.48(1)$ & 
\multirow{2}{*}{\cite{Cohen:2019zev}} \\ 
$R_{J/\psi}^\mathrm{\,opt}$  & $1.11(4)$   & $1.10(3)$ & $1.06(2)$&  \\ \hline
\end{tabular}
\caption{ \sl SM predictions for $R_{J/\psi}^\mathrm{cut}$ and $R_{J/\psi}^\mathrm{opt}$ defined in Eq.~\eqref{eq:RV-cut} and \eqref{eq:RV-opt}, respectively, for different values of $q^2_\mathrm{min}$.}
\label{tab:RJPsi} 
\end{table}

A detailed differential comparison of  
$R_{J/\psi}^\mathrm{cut}(q^2_{\rm min})$
and $R_{J/\psi}^\mathrm{opt}(q^2_{\rm min})$, as a function of $q^2_\mathrm{min}$,
is shown in Fig.~\ref{fig:Bc-JPsi-prediction}.
As can be seen from the right panel, the difference among the two observables is 
more pronounced for small $q^2_\mathrm{min}$ values, while they become
almost equivalent at large $q^2_\mathrm{min}$ values.

\

\paragraph{\underline{$B\to D^\ast\ell \bar{\nu}$.}} 
Despite the form factor uncertainty in $B\to D^\ast\ell \bar{\nu}$ is quite 
small~\cite{Fajfer:2012vx,Grinstein:2017nlq,Bigi:2016mdz,Bigi:2017jbd,Jaiswal:2017rve,Bernlochner:2017jka,Bordone:2019vic,Iguro:2020cpg}, in view of future high-statistics 
measurements it is worth analysing the impact of the improved ratios also in this case.
Here the SM prediction can be obtained by using the shapes of the $A_1$, $A_2$ and $V$ form-factors that are constrained experimentally in the CLN parameterization~\cite{Amhis:2019ckw},  combined with the estimate of the ratio $A_0(q^2)/A_1(q^2)$ obtained in~\cite{Bernlochner:2017jka} 
to which we assign a conservative $10\%$ error. With these inputs, we obtain $R_{D^\ast}^\mathrm{SM}=0.252(5)$ with a~$\approx 2\%$ uncertainty. As shown in Table~\ref{tab:Rall}, this error can be halved  by using the improved LFU ratios 
with a $q^2$ cut at $7~{\rm GeV}^2$ that, similarly to the $B_c\to J/\psi \tau \bar{\nu}$ case,
would retain a large fraction of the signal.
At this level of accuracy, QED corrections, that so far we have neglected, could be become a relevant 
source of uncertainty.

\

\paragraph{\underline{$B_s \to D_s^\ast\ell \bar{\nu}$.}} 
Proceeding in a similar manner we study the $B_s\to D_s^\ast$ transition~\cite{Bordone:2019guc}. Form-factor uncertainties are sizable in this case since lattice QCD results are not yet available at nonzero recoil~\cite{McLean:2019sds}. We consider the  conservative
form-factor estimate in Ref.~\cite{Cohen:2019zev}, from which we obtain the prediction $R_{D_s^\ast}=0.20(2)$ for the standard definition. 
Using the same form factors, we obtain the predictions for the improved observables shown in Table~\ref{tab:Rall}. Already at $q^2_\mathrm{min}=m_\tau^2$, we see that the uncertainty drops to $\approx 4\%$ and 
$\approx 3\%$ for $R_{D_s^\ast}^\mathrm{cut}$ and $R_{D_s^\ast}^\mathrm{opt}$, respectively, which becomes even smaller as $q^2_\mathrm{min}$ increases.

\

\paragraph{\underline{$B_c\to \psi(2S) \ell \bar{\nu}$.}} 
As a final example we discuss the $B_c\to \psi(2S) \ell \bar{\nu}$ case, which might represent an interesting channel at 
hadron colliders. Here no precise estimates of the form factors exist at present. While this fact prevents obtaining precise predictions of the standard LFU ratio,
we can still obtain quite reliable predictions for the improved ratios under rather conservative assumptions. In particular, we employ 
the form factor estimated in Ref.~\cite{Cohen:2019zev} for the $B_c\to J/\psi$ case, replacing the mass [$m_{J/\psi} \to m_{\psi(2S)}$]
and doubling all the errors. Doing so, we obtain the values shown in Table~\ref{tab:Rall}. Given the smaller $q^2$ range
($q^2_{\rm max} \approx 6.8~{\rm GeV}^2$) here we only quote the ratios up to $q^2_{\rm min} =5~{\rm GeV}^2$. There we reach 
a $\approx 3\%$ error on both improved LFU ratios, which is quite remarkable given the large inputs errors.

\begin{table}[t]
\renewcommand{\arraystretch}{1.8}
\centering
\begin{tabular}{|c|ccc|c|}
\hline 
$\,q^2_\mathrm{min}\,$ & $m_\tau^2$ & $\,5~\mathrm{GeV}^2\,$  & $\,7~\mathrm{GeV}^2\,$  & form factors\\ \hline\hline 
$R_{D^\ast}^\mathrm{\,cut}$  & $0.343(7)$ & $0.429(8)$ & $0.496(6)$ &  \cite{Amhis:2019ckw,Bernlochner:2017jka} \\ 
$R_{D^\ast}^\mathrm{\,opt}$  &   $1.11(2)$ & $1.09(2)$ & $1.06(1)$ & (see text) \\ \hline 
$R_{D_s^\ast}^\mathrm{\,cut}$  & $0.29(1)$ & $0.378(8)$ & $0.451(5)$ &  \multirow{2}{*}{ \cite{Cohen:2019zev}\ } \\ 
$R_{D_s^\ast}^\mathrm{\,opt}$  &   $1.09(3)$ & $1.07(2)$ & $1.04(1)$ &  \\ \hline 
$R_{\psi(2S)}^\mathrm{\,cut}$  & $0.16(1)$ & $0.27(1)$ & -- &  \cite{Cohen:2019zev}  \\ 
$R_{\psi(2S)}^\mathrm{\,opt}$  &   $1.10(6)$ & $1.06(4)$ & -- &  (see text) \\ \hline 
\end{tabular}
\caption{ \sl Predictions for the $R_{V}^\mathrm{cut}$ and $R_{V}^\mathrm{opt}$ ratios in different $B\to V \ell \nu$ modes, 
for different values of $q^2_\mathrm{min}$.}
\label{tab:Rall} 
\end{table}

\section{Discussion}\label{sec:conclusion}
The examples presented above provide a clear illustration of the virtues of the improved LFU ratios in obtaining SM predictions 
with a reduced theoretical error, both for cases where the error on the standard ratio is small, 
such as $R_{D^\ast}$, as well as in cases where this error is very large, such as $R_{\psi(2S)}$.
In this section we address three points which might appear more problematic compared to the standard case, 
namely the impact of QED corrections, the experimental error, and the sensitivity to physics beyond the SM.

\

\paragraph{\underline{QED corrections.}}
QED corrections do represent an additional source of LFU breaking within the SM. 
If not properly corrected for, the effects of soft and collinear radiation 
can become relevant in light-lepton decays being of $O[ \alpha \log(m_\mu/m_B)]$ (see e.g.~\cite{Bordone:2016gaq,deBoer:2018ipi}). 
Such collinear logs vanish for inclusive measurements. However, they also vanish at the differential level in the 
$q_0^2$ spectrum~\cite{Bordone:2016gaq,Isidori:2020acz}, where 
\be
q^2_0 \equiv (p_B-p_V)^2~,
\ee
which does not coincide with the dilepton invariant mass spectrum in presence of QED radiation.
Hence we do not expect any specific problem in the extraction of the improved LFU ratios,
as far as QED corrections are concerned, provided they are defined in terms of 
$q_0^2$ rather than $q^2$. 

\

\paragraph{\underline{Experimental accessibility.}}
The need of a differential measurement makes the experimental extraction of the  
improved LFU ratios potentially more challenging at hadron colliders. 
However, some information on the $q_0^2$ distribution is partially available also in these experimental setup,
via the effective determination of the $B$ meson momentum  (see e.g.~\cite{Aaij:2020hsi,Aaij:2017deq}).
Actually an effective lower cut on $q_0^2$  is unavoidable in the busy environment of hadron colliders in order to 
reduce the background of $B \to X \ell \nu$, where $X (\to V)$ is an excited hadronic state of higher mass. 
As a result, we do not expect a significant increase of the error, at least for the extraction of $R_V^\mathrm{\,cut}$, and maybe even 
an advantage given no extrapolation of the signal in a background-dominated region is necessary. 

\

\begin{figure}[t]
  \centering
    \includegraphics[width=.90\columnwidth]{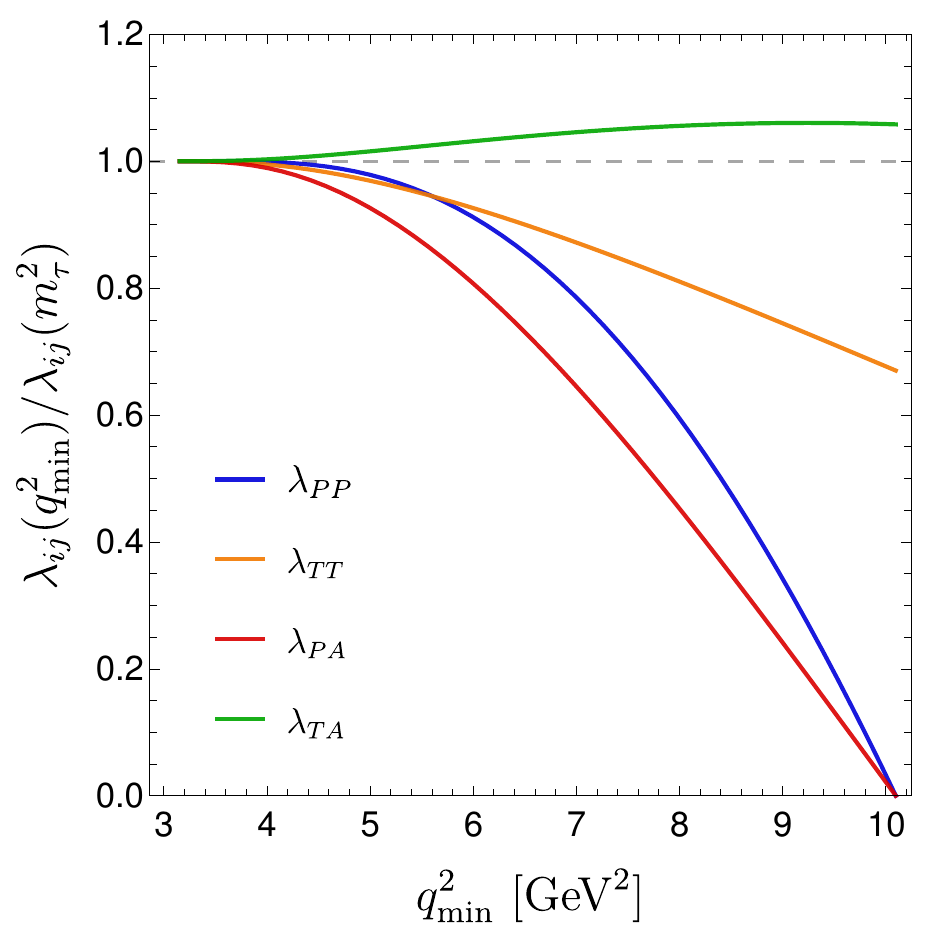}
  \caption{ \sl \small Central values of the $\lambda_{XY}$ coefficients controlling NP effects in  $R_V^{\mathrm{opt}}$, according to  
(\ref{eq:lambdaXY}),  in the     $B_c\to J/\psi$  case.}
  \label{fig:NP}
\end{figure}

\paragraph{\underline{Sensitivity to physics beyond the SM.}}

So far we focused  on the SM predictions of the optimized LFU ratios, which are
the key ingredients to establish a possible deviation from the SM in the comparison with data.
If this is established, the interpretation of the result is modified compared to the non-optimized case,
given the different weights of vector and scalar amplitudes in $R_{V}^\mathrm{cut}$ and $R_{V}^\mathrm{opt}$
compared to $R_V$.
The precise impact needs to be evaluated case by case, however, some general conclusions can be drawn.
To illustrate this point, we consider the most general dimension-six effective Lagrangian encoding SM and 
New Physics (NP) contributions to $b\to c \ell \nu$ transitions (renormalized at a scale $\mu \sim m_b$):
\begin{align}
\mathcal{L}_\mathrm{eff} &= -2\sqrt{2} G_F V_{cb} \Big{[}(1+g_{V_L})\,(\bar{c}_L\gamma_\mu b_L)(\bar{\ell}_L \gamma^\mu \nu_L)\\[0.2em]
&+g_{V_R}\,(\bar{c}_R\gamma_\mu b_R)(\bar{\ell}_L \gamma^\mu \nu_L)+g_{S_R}\,(\bar{c}_L b_R)(\bar{\ell}_R \nu_L)\nonumber\\[0.2em]
&+g_{S_L}\,(\bar{c}_R b_L)(\bar{\ell}_R \nu_L)+g_{T}\,(\bar{c}_R  \sigma_{\mu\nu}b_L)(\bar{\ell}_R \sigma^{\mu\nu}\nu_L)\Big{]}+\mathrm{h.c.}\,.\nonumber
\end{align}
The combinations of effective couplings appearing in $P\to V\ell\bar{\nu}$ decays at the tree level are 
$g_A= g_{V_R} -  g_{V_L}-1$, $g_P=g_{S_R}-g_{S_L}$,  and $g_T$, with the SM case corresponding to 
$g_A=1$ and $g_S=g_T=0$. Normalizing the  optimized ratio to the SM 
expectation we can write 
\begin{eqnarray}
&& \frac{ R_V^{\mathrm{opt}}  \big{(}q^2_\mathrm{min}\big{)}}{  \left. R_V^{\mathrm{opt}}  \big{(}q^2_\mathrm{min}\big{)}\right._{\rm SM} } =
\left| g_{A}  \right|^2 \times \left[ 1  
      + \lambda_{PP} \left| \frac{g_P}{g_{A}}  \right|^2 + \lambda_{TT} \left| \frac{g_T}{g_{A}}  \right|^2  \right. \nonumber \\
&& \qquad\qquad  \left.   +\lambda_{PA}\ \Re\left( \frac{g_P}{g_{A}} \right)   +  \lambda_{TA}\ \Re\left( \frac{g_T}{g_{A}} \right)   
\right]~, \label{eq:lambdaXY}
\end{eqnarray}
where $\lambda_{XY}$ are numerical coefficients which vary according to $q^2_{\rm min}$.
First of all, it is worth noting that a change of $g_A$ (which is one of the most interesting possibility 
according to recent combined analyses of $R_D$ and $R_{D^*}$) 
leads to a breaking of universality which is the same for optimized and non-optimized observables.
Concerning the other type of NP effects, in Fig.~\ref{fig:NP} we report the central values of the  
$\lambda_{XY}$ as a function of $q^2_{\rm min}$ in the $B_c\to J/\psi \ell \bar{\nu}$ case
(using the SM form factors from \cite{Cohen:2019zev} and the tensor one from  Ref.~\cite{Leljak:2019eyw}).
As expected, the sensitivity to scalar amplitudes vanishes for increasing $q^2_{\rm min}$: 
this is  an unavoidable feature of the optimized ratio which, by construction, tends to 
suppress the contribution of the scalar form factor. On the other hand, the sensitivity to 
tensor amplitudes is not significantly affected. 
The optimized observables can thus be considered very clean and sensitive 
probes of possible non-universal effects associated to vector- or tensor-type interactions.

\

In conclusion, we believe the improved observables we have proposed  in this letter do 
represent a valuable tool to reduce the overall error 
of theoretical origin in a wide class of $P\to V\ell \bar{\nu}$ decays.
Their measurement could shed some light on the hints of 
 LFU violations in charged-current interactions.

\section*{Added note}

During the completion of this work,  a new lattice estimate of the $B_c\to J/\psi$ form factors appeared~\cite{Harrison:2020gvo}.
The results in~\cite{Harrison:2020gvo} are perfectly compatible with those in~\cite{Cohen:2019zev} that we have adopted in our numerical analysis, 
but have significantly smaller errors. These new results diminish the need of improved LFU ratios in $B_c\to J/\psi \ell\nu$; however, 
similarly to the $B \to D^* \ell\nu$ case, the observables we have proposed 
can still be used as an independent method to reduce and crosscheck the overall error of theoretical origin.  
In this perspective, for illustrative purposes, we find it still useful to use the conservative errors from~\cite{Cohen:2019zev} 
to analyse the power of the method.  For completeness,  we report here the predictions for 
the improved LFU ratios obtained using the $B_c\to J/\psi$ form factors in~\cite{Harrison:2020gvo}:
 \bea
&& R_{J/\psi}^\mathrm{cut}(m_\tau^2) = 0.331(2)~,  \\
&& R_{J/\psi}^\mathrm{opt}(m_\tau^2) = 1.073(4)~.  
 \eea

\appendix 

\section{Form-factors}\label{app:form-factors}

For completeness, we provide our definition for the $P\to V$ form-factors,
\begin{align}
\label{def:FFV}
\langle V(k)&|\bar{c}\gamma^\mu(1-\gamma_5) b|P(p)\rangle = \varepsilon_{\mu\nu\rho\sigma}\varepsilon^{\ast\nu}p^\rho k^\sigma \frac{2 V(q^2)}{M+m}\nonumber\\[0.3em]
&-i \varepsilon_\mu^\ast(m_B+m_{K^\ast})A_1(q^2)\nonumber\\[0.3em]&+i(p+k)_\mu (\varepsilon^\ast \cdot q)\frac{A_2(q^2)}{M+m}\nonumber\\[.3em] 
&+i q_\mu(\varepsilon^\ast \cdot q) \frac{2 m}{q^2}[A_3(q^2)-A_0(q^2)]\,,
\end{align}
\noindent where $2 m A_3(q^2)=(M+m)A_1(q^2)-(M-m)A_2(q^2)$ and we have used the convention $\varepsilon_{0123}=+1$.

\vspace*{1.3em}

\section*{Acknowledgements}

We thank Guenther Dissertori and Yuta Takahashi for asking the questions that gave rise to this work. We are also grateful to
Hank~Lamm for useful correspondence about~\cite{Cohen:2019zev}. This project has received funding from the European Research Council (ERC) under the European Union's Horizon 2020 research and innovation programme under grant agreement 833280 (FLAY), and by the Swiss National Science Foundation (SNF) under contract 200021-175940.

\end{document}